\documentclass[aps,pra,twocolumn,groupedaddress,showpacs]{revtex4}

\def\tphi{\tilde{\phi}}

\def\openone{\leavevmode\hbox{\small1\kern-3.8pt\normalsize1}}
\def\RR{{\rm I\kern-.2emR}}
\def\tr{{\rm tr}\; }

\newcommand{\ket}[1]{| #1 \rangle}
\newcommand{\bra}[1]{\langle #1 |}

\newcommand{\outerp}[2]{\ket{#1}\! \bra{#2}}

\newcommand{\bitem}{\begin{itemize}}
\newcommand{\eitem}{\end{itemize}}
\newcommand{\benum}{\begin{enumerate}}
\newcommand{\eenum}{\end{enumerate}}
\newcommand{\beq}{\begin{equation}}
\newcommand{\eeq}{\end{equation}}
\newcommand{\beqa}{\begin{eqnarray}}
\newcommand{\eeqa}{\end{eqnarray}}
\newtheorem{definition}{Definition}

\newtheorem{proposition}{Proposition}

\newcommand{\bproof}{\begin{proof}}
\newcommand{\eproof}{\end{proof}}
\newcommand{\bprop}{\begin{proposition}}

\newcommand{\bdef}{\begin{definition}}

\begin{document}


\title{Largest separable balls
around the maximally mixed bipartite quantum state}


\author{Leonid Gurvits and Howard Barnum}
\affiliation{ CCS-3, Mail Stop B256, Los Alamos National Laboratory,
Los Alamos, NM 87545} 
\affiliation{}


\date{\today}

\begin{abstract}
For finite-dimensional bipartite quantum systems,
we find the exact size of the largest balls, in spectral $l_p$
norms for $1 \le p \le \infty$, of separable (unentangled)
matrices around the identity matrix.  
This implies a simple and intutively meaningful
geometrical sufficient condition for separability
of bipartite density matrices: 
that their purity $\tr \rho^2$ not be too large.
Theoretical and experimental applications
of these results include algorithmic problems such as computing whether
or not a state is entangled, and practical ones such as obtaining 
information about 
the existence or nature of entanglement in 
states reached by 
NMR quantum computation implementations or other experimental
situations. 
\end{abstract}


\maketitle


\section{Introduction and summary of results}
Entanglement is
an important element of many quantum information processing 
procedures, from cryptography
to computation to quantum 
teleportation.  Indeed, a quantum algorithm
operating on pure quantum states 
must entangle a number of qubits increasing unboundedly with 
the input size, if it is not to be simulable in polynomial 
time on a classical computer \cite{Jozsa2002a}.  It is not
known whether this is so when the computer state may
be mixed.  Determining whether a given state, even of two quantum systems,
is entangled or separable (not entangled)
is in general difficult, and considerable
effort has been expended on finding necesssary and/or sufficient
conditions.  The normalized separable states form a convex set.
A key aspect of the geometry of a convex set is the size of the
largest ball (especially in $l_2$ norm) that fits entirely inside it,
and the smallest ball that covers it.  We here find the inner 
ball for the set of separable 
quantum states.  The result has both practical and theoretical
relevance.  For example, it provides a simple sufficient criterion 
for separability. 
A bipartite state of
a composite system with overall dimension $d$ is separable if 
its purity $\tr \rho^2$ is less than $1/(d-1)$ (as conjectured
in \cite{Zyczkowski98a}).
Because of its simple geometric nature and ease of computation, 
this criterion is likely to 
be very useful both in theoretical applications and in analyzing 
whether entanglement was present in experiments.  
Just as importantly, knowing the size of such balls helps one understand the
computational complexity of problems involving a convex set.  For example,
using bounds on the size of the inner ball (rather than the
exact result we here present) one
of us has shown the NP-hardness of the ``weak membership'' problem 
for separability when the dimensions of the two systems are not
too different \cite{Gurvits2002a};  
it is likely that the exact result reported herein
may be used in extending this hardness result or in obtaining 
other complexity results
about separability and entanglement.

Our main results begin with Theorem 1, that
the matrix $I + \Delta$ is separable for 
all hermitian $\Delta$ with $||\Delta||_2 \le 1$.
Corollary 1 gives similar statements for 
other $p$-norms.  Theorem 3 
establishes that for $2 \le p \le \infty$, 
these are the largest such balls.
A similar result is easy for $1 \le p \le 2$; 
it involves subtracting a normalized pure state to reach the
edge of the ball, where the state becomes singular and thus
is at the edge of the positive cone.
Theorem 4 tells us about the size of the largest negative
eigenvalues of the partial transpose of 
bipartite positive matrices
that are rank-$m$ projectors, giving us information about how quickly
we can hit the entangled matrices when departing from the identity
by adding a positive multiple of 
such a projector.  In particular, by considering a perturbation
$\Delta$ proportional to such a projector whose partial transpose has
maximal modulus of its most negative eigenvalue, 
it can be used to show
that for $p=2$ the largest $p$-norm ball around $I$
touches the edge of the separable cone at places within the positive
cone.  

As one example of practical relevance, the ``pseudopure''
states which describe each molecule 
in NMR quantum information processing are
mixtures of the uniform density matrix with a pure state;  
the signal of
quantum dynamics derives from the small pure component.
Because of this, the density matrices of 
the different nuclear spins in a given molecule
have not, in experiments so far done, exhibited entanglement
despite the pure component being an entangled state; 
they have remained within known lower bounds on the
size of the ball of separable states \cite{Braunstein99a}.  
Our determination of the exact size of this ball for bipartite 
entanglement increases known lower bounds on the polarization
necessary in order for such bulk computation on pseudopure
states to be able to achieve bipartite entanglement, although
due to the bipartite nature of our analysis, it does not rule
out the production of entangled states that are separable 
with respect to every bipartition, at lower polarization.
This raises the interesting question whether the exponential 
gap between our bound for bipartite separability and known
bounds for separability in this context, can be closed.
Other quantum information processing
procedures may also involve such mixtures;  also, 
mixture with the identity matrix
is a frequently studied model of quantum noise, the ``depolarizing
channel,'' to which our 
results are relevant.  When this occurs in bipartite contexts, our
results are much stronger than previously known.  
We emphasize, though, that the sufficient conditions for
entanglement and separability provided by our results apply
in arbitrary contexts, not only for mixtures of a pure state with
the normalized identity.

\section{Mathematical background and notation}
We represent unnormalized states of a
quantum system composed 
of two subsystems of dimensions 
$M$ and $N$ (``$M \otimes N$ system''), 
as positive semidefinite $M \times M$
block matrices, with $N \times N$ blocks (so that
they are $MN \times MN$ complex matrices).  
(These are the elements of the 
unnormalized density
operator, in some fixed basis of product states $e_i \otimes f_j$.)
Rather than Dirac notation, we use roman letters for
vectors, but we use $\dagger$ for
the adjoint. 
Such a matrix $A$ is called {\em separable} if it can
be written 
\beqa
A = \sum_i x_i x_i^\dagger \otimes y_i y_i^\dagger\;.
\eeqa
Objects like $x_i x_j^\dagger$ are outer products
(in Dirac notation
$\outerp{x_i}{x_j}$; here $\ket{x_i}$ are not 
assumed normalized.).  We use $e_i$ for elements of
an orthonormal basis (typically $\ket{i}$ in Dirac notation).
Thus our $e_i \otimes e_j$ would typically be written 
$\ket{e_i} \otimes \ket{e_j}$, 
or $\ket{e_i} \ket{e_j}$ or simply $\ket{i}\ket{j}$ in Dirac notation.  
$M_n$ is the set of $n \times n$ complex matrices, 
$M_{mn}$ the set of $m \times n$ complex matrices.
When interpreting tensor products as block matrices 
the left-hand factor corresponds to ``which block,''
and the right-hand to the indices within blocks.

$||X||$ (or $||X||_{\infty}$),
with a matrix $X$ as argument, is the usual operator
norm induced by Euclidean norm $||x|| = \sqrt{(x,x)}$ on vectors
(i.e. $||X||:= \sup_{||x||=1} ||Xx||)$.  (It is also the $\l_{\infty}$
norm of the vector of singular values of $X$, i.e. the largest
singular value.)  $||X||_1$
is $\tr \sqrt{X^\dagger X},$ the sum of the
singular values of $X$.  $||X||_2$, the Frobenius
norm, is $\sqrt{\tr X^\dagger X}$, the Euclidean norm associated
with the inner product $\tr X^\dagger Y$.
The squared Frobenius norm
is also the sum of squared singular values of $X$, and 
the sum of squared moduli of $X$'s matrix elements.  We write
$[a_{ij}]$ for the matrix with elements $a_{ij}$.

Linear maps $\phi: M_m \rightarrow M_n$
are called {\em positive} if they preserve positive semidefiniteness.
They also preseve Hermiticity  
(write Hermitian $H$ as a sum of positive and negative semidefinite
parts, and use linearity and positivity).
A {\em stochastic} map takes the identity matrix $I$ to itself.  
We may apply such a map $\phi$ to one
subsystem of a bipartite system, while doing nothing to 
the other system.   Applying it to the $N$-dimensional subsystem
is just applying it to each block 
of the block matrix $X$; we call the resulting map on the 
bipartite system $\tilde{\phi}$:
\beqa
\tilde{\phi}(X) := 
\left( \begin{array}{cccc}
		  \phi(X_{1,1}) & \phi(X_{1,2}) & \dots & \phi(X_{1,N})\\
		  \phi(X_{2,1}) & \phi(X_{2,2}) & \dots & \phi(X_{2,M})\\
		  \dots &\dots & \dots & \dots \\
		  \phi(X_{N,1}) & \phi(X_{N,2}) & \dots & \phi(X_{N,N})
\end{array} \right).
\eeqa

An important condition equivalent to separability 
of $A$ is that for any stochastic positive linear map $\phi$,
$\tphi(A)$ be positive semidefinite.  We refer to it as the
``Woronowicz condition.'' 
This appeared in \cite{Horodecki96a}, 
but was already essentially
proven (along with the sufficiency of the partial transposition
map (``Woronowicz-Peres criterion'')
for two qubits or a $2 \otimes 3$ system, and a $2 \otimes 4$ 
counterexample)
in \cite{Woronowicz76a}, though the terminology
of separability and entanglement is not used there.  The proof there is
given for a $ 2 \otimes N$ system ($N < \infty$), 
but it works for $M \otimes N$
by expanding the range of an index.  

We also use the following fact.\\
\noindent
{\em Fact:} Let $\Delta$ be an $M \times M$ block matrix
(whose blocks need not be square).  Define $\Delta'$ as the
matrix whose elements are the operator norms of the blocks of 
$\Delta$.  Then $||\Delta|| \le ||\Delta'|| \le M ||\Delta||$.
The first inequality is well-known;  the second holds because 
$||\Delta_{ij}|| \equiv ||P_i \Delta Q_j|| \le ||\Delta||$.
($P_i$ ($Q_i$)is the projector onto the $i$-th subspace in the 
direct sum decomposition of the row (column) space that defines
the blocks.)  So, by adding to $\Delta'$ a matrix with nonnegative entries
(therefore not decreasing the norm) we
can obtain $||\Delta||$ times the $M\times M$ all-ones matrix, whose
norm is $M$.

\section{Main result:  Separability of perturbations of the identity}

We give two proofs of the main result.  Both proceed via Proposition 
2, which states that stochastic positive linear maps on 
$n \times n$ matrices are contractive
with respect to the $\l_\infty$ (``operator'') norm of matrices, 
for {\em all} matrices.  (The result for Hermitian matrices only is
much easier.)   Those interested only in the shortest proof,
which uses the Naimark extension,
may skip to the statement of Proposition 2 below.  We think it is of
interest to see the connections of the norm contraction result to two
different concepts well-known to quantum information theorists:  in
in the second proof, the
Naimark extension and in the first proof, separability.
The first proof proceeds via Proposition 1, which is a special 
case of recent results by one of us providing sufficient criteria
for separability.

\noindent
{\em Proposition 1:}  If $||X|| \le 1$, the block matrix 
\beq 
\left( \begin{array}{ll}
I & X \\
X^\dagger & I
\end{array}
 \right)
\eeq
is separable.
This is the $M=2$ case of a recent theorem 
\cite{Gurvits2001a} that all positive 
semidefinite $M \times M$ block Toeplitz or block Hankel matrices
whose blocks are $N \times N$ matrices
are separable, whose proof we include here.  
The paper \cite{Gurvits2001a} also
contains two alternative proofs for the
special case $M=2$. One of those proofs was independently
discovered in \cite{Ando2001a}.

\noindent
{\em Proof (of separability of positive semidefinite block
 Toeplitz matrices \cite{Gurvits2001a}):}
The proposition is a corollary to the following Lemma:\\
\noindent
{\it Lemma:}
Consider an $((M+1) \times N)$  positive semidefinite 
block Toeplitz matrix $T$:
$$
T = \left( \begin{array}{ccccc}
		  R_0 &  R_1 & R_2 &\dots & R_M\\
		  R_1^{\dagger} & R_0 & R_1 & \dots & R_{M-1}\\
		  R_2^{\dagger}  & R_1^{\dagger} & R_0 & \dots & \dots \\
		  \dots  & \dots & \dots & \dots & \dots \\
		  R_M^{\dagger} & R_{M-1}^{\dagger} & R_{M-2}^{\dagger} & \dots & R_{0}\end{array} \right).
$$
(This structure is the definition of a Toeplitz matrix.)
Suppose that $Rank(T) = K$. Then there exist an $N \times K$ matrix $X$ 
and a $K \times K$ unitary
matrix $U$ such the $T(i,j) = X U^{i-j} X^{\dagger}, 0 \leq i,j \leq M-1$. \\
{\em Proof:} 
Since our matrix $T$ is positive
semidefinite with $Rank(T) = K$, $T = Y Y^{\dagger}$, where
$$
Y = \left( \begin{array}{ccccc}
		  Y_0 &
		  Y_1 &
		  Y_2 &
		  \dots &
                  Y_M \end{array} \right)^T ,
$$
and each block $Y_{i}$ is an $N \times K$ matrix. Define 
the upper submatrix $Y_{U}$ as
$$
Y_{U} = \left( \begin{array}{ccccc}
		  Y_0 &
		  Y_1 &
		  Y_2 &
		  \dots   &
                  Y_{M-1} \end{array} \right)^T ,
$$
and, correspondingly, the lower submatrix $Y_{L}$ as
$$
Y_{L} = \left( \begin{array}{ccccc}
		  Y_1 &
		  Y_2 &
		  Y_3 &
		  \dots   &
                  Y_M \end{array} \right)^T.
$$
It follows straight from the Toeplitz structure that 
$Y_{U} Y_{U}^{\dagger} = Y_{L} Y_{L}^{\dagger}$.  Thus 
there exists an unitary $K \times K$ matrix $U$ 
such that $Y_{L} = Y_{U} U$ or in other words :
$$
Y = \left( \begin{array}{ccccc}
		  Y_0 &
		  Y_{0}U &
		  Y_{0}U^{2} &
		  \dots   &
                  Y_{0} U^{M-1}\end{array} \right)^T.
$$
Recalling that $T = Y Y^{\dagger}$ , we finally get the identities
$$
T(i,j) =X U^{i-j} X^{\dagger} , 0 \leq i,j \leq M-1 ; X = Y_0.
$$
\noindent
{\em Corollary:}
Using the notation of the proof above, 
put $U = V Diag (z_1,...,z_K) V^{\dagger}$ where $V$ is unitary
and the complex numbers $z_i$ have norm one , i.e. 
$\overline{z_{i}} = z_{i}^{-1} , 1 \leq i \leq K$.
Denote the $i$th column of $XV$ as $L_i$  and \\
$Z_i  = (1,z_{i},...,z_{i}^{M-1})^{T}$ , $1 \leq i \leq K$.
Then the following ``separability'' representation holds :
$$
T =  \sum_{1 \leq i \leq K} Z_{i} Z_{i}^{\dagger} 
\otimes L_{i} L_{i}^{\dagger}.
$$

We can use Proposition 1 to show a contraction inequality (Proposition
2 below).  We got the idea of using separability to obtain
operator inequalities involving stochastic positive maps from
\cite{Woronowicz76a}, where the Kadison inequality $\phi(X^{2}) \ge
(\phi(X))^{2}$ for stochastic $\phi$ and hermitian $X$ is implicitly
connected with the separability of:
$$
\rho = \left( \begin{array}{ll}
I & X \\
X & X^{2}
\end{array}
 \right)\;.
$$
In a sense, for $M=2$  separability of $\rho$ is equivalent to 
Proposition 1, as in this case  there is a local unitary transformation
$\rho \mapsto (A \otimes I)  \rho (A^{\dagger} \otimes I)$ which maps block 
Toeplitz matrices to block Hankel ones (see \cite{Gurvits2001a} ) :
$$
A  = \frac{1}{\sqrt{2}} \left( \begin{array}{ll}
1& i \\
i & 1
\end{array}
 \right)\;.
$$
One can probably prove the next proposition 
using the Kadison inequality and this transformation.

\noindent {\em Proposition 2:}
Let $\phi: M_n \rightarrow M_n$ be a stochastic positive linear map.
Then for any $X\in M_n$, $||\phi(X)|| \le ||X||.$

\noindent
{\em Proof 1:}
We show that $\phi(X) \le 1$ if $||X|| \le 1$; the proposition follows
by $\phi$'s linearity.  Apply $\tphi$ to the separable
state
of Proposition 1, obtaining:
\beq \label{joe}
\left( \begin{array}{ll}
I & \phi(X) \\
\phi(X^\dagger) & I
\end{array}
 \right)\;.
\eeq

Write $X = X_1 + iX_2$ with $X_1, X_2$ hermitian.
Then $\phi(X^\dagger) = \phi(X_1^\dagger - i X_2^\dagger) 
= $ 
(by Hermiticity preservation of $\phi$, 
which follows from its 
positivity)
$\phi(X_1)^\dagger - i \phi(X_2)^\dagger
= \phi(X_1 + i X_2)^\dagger = \phi(X)^\dagger$.
Hence (\ref{joe}) is equal to:
\beq 
\left( \begin{array}{ll}
I & \phi(X) \\
\phi(X)^\dagger & I
\end{array}
 \right)\;.
\eeq
Since this resulted from applying $\tphi$ to a 
separable state, it is a positive semidefinite matrix.
Positivity of this matrix is equivalent 
(cf. e.g. \cite{Horn85a}, p. 472)
to $\phi(X)^\dagger \phi(X) \le I$; {\em i.e.}, 
$x^\dagger \phi(X)^\dagger \phi(X) x \le 1$ for all normalized
$x$, i.e. $||X||\le 1$.  
 
 This proof was independently discovered in \cite{Ando2001a}. Let us
present a very different proof which does not use separability but
another concept well known in the quantum information community.

\noindent
{\em Proof 2 (lifting) :}
It is well known that the extreme points of the matrix ball
$\{X : ||X|| \le 1 \}$ are unitary matrices. 
Thus we can assume that $X$ is unitary , i.e. 
$X =\sum_{1 \le i \le N} z_{i} e_{i}e_{i}^{\dagger}$ , where $|z_{i}| = 1 , 1 \le i \le N$  and
$\{e_{i} , 1 \le i \le N \}$ is an orthonormal basis in $C^{N}$.
Thus $\phi(X) = \sum_{1 \le i \le N} z_{i} Q_{i} ,  Q_{i} = \phi(e_{i}e_{i}^{\dagger}).$
Since $ \phi$ is a positive stochastic map,
$$
Q_{i} \ge 0  (1 \le i \le N ) \mbox{ and } I = \sum_{1 \le i \le N}  Q_{i}.
$$
By Naimark's theorem \cite{Naimark40a}
\cite{Naimark43a} 
(cf. \cite{Peres93a} for a simple exposition in finite
dimension) 
there exist commuting orthogonal projectors $P_{i} : C^{K}
\rightarrow C^{K}, N \le K \le N^{2},$ and a unitary injection $U :
C^{N} \rightarrow C^{K}$ such that
$$
Q_{i} = U^{\dagger} P_{i} U (1 \le i \le N ) \mbox{ and } I = \sum_{1 \le i \le N}  P_{i}.
$$
It is easy to see that $ ||\sum_{1 \le i \le N} z_{i} P_{i}|| \le 1 $.
Thus 
\beqa
||\sum_{1 \le i \le N} z_{i} Q_{i}||  \le  ||U^{\dagger}|| ||\sum_{1 \le i \le N} z_{i} P_{i}|| ||U|| \le 1.
\eeqa
This second proof suggests that there
might be a deeper connection between Naimark's theorem  and separability.

We proceed to the main theorems.

\noindent
{\em Theorem 1:}
The matrix $I + \Delta$ is separable for 
all hermitian $\Delta$ with $||\Delta||_2 \le 1$.

\noindent
{\em Proof:}  \
\beqa
||\tphi(\Delta)||^2 \le ||A||^2 \le ||A||^2_2,
\eeqa
where $A := [a_{ij}]$, $a_{ij} := ||\phi(\Delta_{ij})||$.
\beqa
||A||_2^2 = \sum_{ij} a_{ij}^2 = \sum_{ij} ||\phi(\Delta_{ij})||^2.
\eeqa
(The first inequality is because the operator norm of a block matrix
is bounded above by that of the matrix whose elements are the norms
of the blocks, and the second is because the Frobenius norm is an upper
bound to the operator norm.)
But
$||\phi(\Delta_{ij})||^2 \le ||\Delta_{ij}||$ by Prop. 2, and this in
turn is less than $||\Delta_{ij}||_2$.  
So
\beqa
||\tphi(\Delta)||^{2}  \le\sum_{ij} ||\phi(\Delta_{ij})||^2 \le \sum_{ij} ||\Delta_{ij}||^2_2 
\equiv ||\Delta||^2_2 \le 1,
\eeqa
the last inequality being the premise of the theorem.
Having shown that $||\tphi(\Delta)|| \le I$, and also 
using $\tphi(I)=I$, we get $\tphi(I + \Delta) \ge 0$, so that
by Woronowicz' criterion 
$I + \Delta$ is separable.
 
\section{Corollaries and additional results:  Maximality of balls, scaling and specific perturbations}
Let us now present some corollaries of Theorem 1. 
Define $||\Delta||_{p} :=
(\sum_i  |\lambda_i| ^{p})^{\frac{1}{p}}$ ($\lambda_i$ being the 
eigenvalues
of the square hermitian matrix $\Delta$.) \\
\noindent
{\em Corollary 1 ($l_{p}$ balls):} Consider an $N \times N$ system.Then
the matrix $I + \Delta$ is separable for all hermitian $\Delta$ with
$||\Delta||_{p} \le 1 (1 \le p \le 2 )$ and $||\Delta||_{p}
\le B(N,p) =: N ^{\frac{2}{p}-1} (2 \le p \le \infty )$.\\
\noindent
{\em Proof:} \ The statements follow from basic $p$-norm inequalities:
the first from the $q=2$ cases of 
$||\Delta||_p \ge ||\Delta||_q$ (for $1 \le p \le q$), the
second from the $q=2$ cases of  
$||\Delta||_q \le n^{\frac{1}{q} -\frac{1}{p}} ||\Delta||_p$ (for $p \ge q$). 
Note that the
dimension $n$ is $N^2$ in our case.  (These inequalities are equivalent
to similar ones for the vector $p$-norms 
$||x||_p := (\sum_i x_i^p)^{\frac{1}{p}}$;  the first set can be 
proved by changing variables to $y_i = x_i^p$ and
using the triangle inequality for norms, the second by letting 
$y_i = x_i^q$ and using the convexity of $f: z \mapsto z^\alpha$
for $\alpha \ge 1$ ($\alpha = p/q$ in our case).)
\\
The $l_{\infty}$ result was also 
obtained very recently in \cite{Ando2001a}
using quite different methods.  The $p$-balls in Cor. 1 are clearly
the largest possible for $1 \le p \le 2$:  by subtracting any normalized
pure state, 
for which all these $p$-norms are $1$, we can leave the positive, 
and hence the separable, cone.  What about $2 < p \le \infty$?  Theorem
3 will show that these are the largest balls for these norms,
too.

The next theorem gives information about how fast
we can reach the entangled states by 
perturbing the identity in a specific direction:  adding a positive multiple
of a pure state.  The main point is that, perhaps surprisingly, the entangled
states are reached fastest by perturbing with a $2 \otimes 2$ Bell state,
rather than, say, a maximally entangled state.
\\
\noindent
{\em Theorem 2 (Perturbation by positive multiples of pure states):}  
1. Consider a pure $\rho$ corresponding to
a state $\ket{\psi} = \sum_{ij} \psi_{ij} e_i \otimes e_j$, 
$1 \le i,j \le N \}.$
The the spectrum of $\rho^{T}$ is 
$(d_{1}^{2} , ..., d_{N}^{2} ; d_{i} d_{j} , -d_{i} d_{j} ( 1 \le i
\neq j \le N ) )$, where $d_{1},...,d_{N}$ are the singular values of 
the $N \times N$ matrix  $\psi := [\psi_{ij} : 1 \le i,j \le N]$
(thus $d_{1}^{2} + ... + d_{N}^{2} = 1 )$.\\
2. Define
$$
W(N) = \min_{\rho \in Den(N,N)} - \lambda_{min}  (\rho^{T}) ,
$$
where $Den(N,N)$ is the set of density 
matrices of $N \times N$ systems. Then $W(N) =
\frac{1}{2}$. \\ 3. If $I + a\rho$ is separable for all $\rho$ and $a
> 0$ then $ a \le W(N)^{-1}= 2.$ \\
\noindent
{\em Proof:} Diagonalize $\psi$ by local unitaries using
the singular value (``Schmidt'') decomposition 
obtaining $Diag(d_1 , ..., d_N).$ The corresponding density matrix has
blocks $\rho_{ij} = d_i d_j e_i e_j^\dagger$, and the spectrum of
$\rho^T$ (which is not changed by applying local unitaries prior to
partial transposition) is as given in Part 1 of Thm.2.
The bound in Part 2 follows from $2ab \le a^2 + b^2$ and is
achieved by $(1/\sqrt{2}, 1/\sqrt{{2}},0,...,0)$. 
The Woronowicz-Peres (WP) condition gives Part 3.

Contrary to the ``folklore,'' in the result above the fully entangled
state is not the worst one; rather, the worst is a maximally
entangled state of two local two-dimensional subspaces.

Theorem 3 establishes the maximum size of the
$p$-balls for $p>2$.  The proof involves 
considering perturbations by a positive multiple of the maximally 
entangled state and establishes when this procedure hits
the entangled states.  
Before formulating the theorem we introduce some notation.
$WP(N,N)$ is the closed convex cone of $N^2 \times N^2$ 
positive matrices satisfying the WP condition, 
i.e. $\rho
\in WP(N,N)$ iff $\rho \ge 0$ and $\rho^{T} \ge 0$.  $Sep(N,N)$
is the closed convex cone of {\em separable} positive
matrices. Obviously $WP(N,N) \subset Sep(N,N)$.  Both
$WP(N,N)$ and $Sep(N,N)$ are 
subsets of the real $N^4$-dimensional linear space
$H(N^2)$ of hermitian $N^2 \times N^2 $ matrices.  The cone 
dual to a convex set $X$ (which need not be a cone) is 
$X^{*} 
:= \{y : \langle y,x \rangle \geq 0 , \forall x \in X \} .$ \\
\noindent
{\em Theorem 3:} Suppose $p > 2$.  If the $p$-ball $Ball(N,p,a)
= \{A \in H(N^2) : A = I + \Delta$, $||\Delta||_{p} \le a \}$ belongs
to $WP(N,N)$ then $ a \le B(N,p) =: N ^{-1 + \frac{2}{p}} (2 \le p \le
\infty )$.

As $WP(N,N) \subset Sep(N,N)$ this Theorem proves that the
$l_{p}$-balls $Ball(N,p, B(N,p))$ in Corollary 1  are largest possible.

\noindent
{\em Proof of Theorem 3:} It is easy to see that the cone dual to $Ball(N,p,a)$, i.e.
$Ball(N,p,a)^{*}$, is $\{ A \in H(N^2) : tr(A) \geq a ||A||_{q} , q =
\frac{p}{p-1} \}$.  It is known \cite{Woronowicz76a} that $WP(N,N)^{*} =
\{\rho_{1} + \rho_{2}^{T} : \rho_{i} \ge 0 , i =1,2 \}$.  If
$Ball(N,p,a) \subset WP(N,N)$ then $WP(N,N)^{*} \subset Ball(N,p,a)^{*}$
and at least $tr(\rho) = tr(\rho^{T}) \geq a ||\rho^{T}||_{q} $ for all
$\rho {\ge} 0 $.  Consider the fully entangled pure $N \times N$
state $\rho_{E}$. Then $tr(\rho_{E})= 1$ . It follows from Part 1 of
Thm. 2 that $\rho_{E}^{T}$ has $N +\frac{N(N-1)}{2}$ eigenvalues equal
to $\frac{1}{N}$ and $\frac{N(N-1)}{2}$ eigenvalues equal to
$\frac{-1}{N}$.  Thus we get that $1 \geq a ||\rho_{E}||_{q} = a
N^{\frac{2-q}{q}}$ and, finally, $a \le N^{\frac{q-2}{q}} =
N ^{\frac{2}{p}-1}$.

The following corollary of Theorem 1 gives (as is evident from the 
proof) the strongest sufficient condition for separability 
of $A \ge 0$ that can be derived by scaling
(considering all ways of writing $A = \zeta ( I + \Delta)$ with 
$\zeta > 0$) and using the Frobenius norm case of Theorem 1 
(applied to $\Delta$).   

\noindent {\em Corollary 2 
(scaling) :} Let $A$ be an (unnormalized)
density matrix of a bipartite system with total dimension $d = NM$ and
$\lambda =(\lambda_{1},...,\lambda_{d})$ be the vector of eigenvalues
of $A$. If \beqa \label{eq: scaling} S(\lambda) = : d -
\frac{||\lambda||_{1}^{2}}{||\lambda||_{2}^{2}} \le 1 \eeqa then $A$
is separable.

\noindent
{\em Proof:} \ It is easy to see that $S(\lambda) = \min_{a > 0} || a
\lambda -e||_{2}^{2}$ , where $e$ is a vector of all ones.  Therefore
if $S(\lambda) \le 1 $ then $A =b (I + \Delta)$ , where $b > 0$ and
$||\Delta||^2_2 \le 1$.  It follows from Theorem 1 that $A = b (I +
\Delta)$ is separable.

\noindent
{\em Corollary 3 (largest Frobenius ball for density matrices):}
Suppose that $A$ is a normalized density matrix of a bipartite
system with total dimension $d = NM$, i.e. $\sum_{1 \le i \le d}
\lambda_{i} = 1$ and $ \lambda_{i} \geq 0 , 1 \le i \le d $.  
If $||A -\frac{1}{d} I ||^2_2 = ||\lambda
-\frac{1}{d}e||^2_2 \le \frac{1}{d(d-1)} = : r^{2}$ then $A$ is
separable.  $r$ is the largest such constant.

\noindent
{\em Proof:} \ Define $t = \lambda -\frac{1}{d}e$.  Then $
||\lambda||_{1}^{2} = 1$ and $||\lambda||_{2}^{2} = \frac{1}{d} +
||t||_{2}^{2} $. Thus $ S(\lambda) = d - \frac{1}{\frac{1}{d} +
||t||_{2}^{2}}$ and $ S(\lambda) \le 1$ iff $ ||t||_{2}^{2} \le
\frac{1}{d(d-1)}$.  From Corollary 2 it follows that $A$ is
separable.  On the other hand $r = 1/\sqrt{d(d-1)}$ is the
radius of the largest ball inside the $d$-dimensional simplex.

\noindent
{\em Remark:} In terms of the ``purity'' $\tr \rho^2$ of the density
matrix (which takes the value $1$ for pure states and $1/d$ for the
maximally mixed state), Corollary 3 says that $\rho$ is separable if
its purity is less than or equal to $1/(d-1)$.

One might conjecture that for $N \times N$ bipartite systems (so $d =
N^2$), any $\lambda$ not satisfying (\ref{eq: scaling}) is the
spectrum of some non-separable positive matrix.  This is not so:  
a sufficient condition for
separability of two-qubit density matrices in terms of the spectrum
is \cite{Verstraete2001a} $\lambda_1 - \lambda_3 - \sqrt{\lambda_2
\lambda_4} \le 0$, where $\lambda_i$ are decreasingly ordered.  This
can hold when the purity is greater than $1/3$, as also noted in 
\cite{Zyczkowski2001a}.

\noindent
{\em Corollary 4:} The matrix $I +a \rho$, where $\rho$ is an $N \times N$
state, is separable if $-1 \le a \le \frac{N^{2}}{N^{2}-2}.$ \\
\noindent
{\em Proof:} \ Clearly it is enough to prove this for pure states.  In this
case the vector of eigenvalues of $I +a \rho$ is
$$
\lambda_{a} =: (1,1,...,1,1+a).
$$
Direct computation gives that $S(\lambda_{a}) = N^{2} -\frac{( N^{2} +
a)^{2}}{N^{2} + 2a + a^{2}}.$ \\ It follows that $S(\lambda_{a}) \le 1
$ iff $ -1 \le a \le \frac{N^{2}}{N^{2}-2}.$ \\

\noindent
{\em Corollary 5:}
If we consider the normalized mixtures
$\sigma = (1 - \epsilon) I/d + \epsilon \rho$,
for pure $\rho$, and scale them as 
\beq
\sigma = \frac{1 - \epsilon}{d}(I + \frac{d \epsilon}{1 - \epsilon} \rho)\;,
\eeq
by Corollary 4 these are separable if $\epsilon \le 1/(d-1) \equiv 1/(N^2-1)$.

A very slightly better, but messier, 
bound can be obtained by solving a quadratic equation
derived from Cor. 2, reminding us 
that the most obvious or tractable scaling is not 
generally the best.  Corollary 5 is of course also true for mixed
$\rho$. 

\section{Perturbation of the identity by positive multiples of projectors}
\label{sec: adding projectors}
A final result again illustrates the
power of scaling (Corollary 2).  It gives us the most 
negative eigenvalue of a partial transpose of
a projector on a bipartite system.  This is interesting because
it tells us when we will
hit the entangled matrices if we add a positive multiple of
that projector to the identity.  Define
$$
W_{m}(N) := \min_{\rho \in PR(m,N)} - \lambda_{min}  (\rho^{T}) ,
$$
where $PR(m,N)$ stands for the compact set of all rank $m$ orthogonal
projectors in $C^{N} \otimes C^{N}$.  Notice that part 2 of
Theorem 2 states that $W_{1}(N) = W(N) = \frac{1}{2}$ ; clearly
$W_{N}(N) = 0$ , it follows from Theorem 1 that also $W_{N-1}(N) = 0$
; it is easy to prove that if $\frac{K}{L}$ is an integer then
$\frac{W_{K}(N)}{K} \le \frac{W_{L}(N)}{L}$. \\ {\em Theorem 4 :}
$$
W_{\frac{N(N-1)}{2}}(N) = \frac{N-1}{2}.
$$
\noindent
\noindent
{\em Proof:} \ Let us define the following operator intervals :
$$
Int(a,N) = : \{ \rho : C^{N} \otimes C^{N} \rightarrow C^{N} \otimes
C^{N} \ : (1+a) I \ge \rho \ge I \}
$$
It follows by a straightforward rescaling from the $l_{\infty}$ part
of Corollary 1 and Theorem 3 that
\begin{description}
\item[Property 1] if $0 \le a \le \frac{2}{N-1}$ then all matrices in
$Int(a,N)$ are separable (and thus satisfy the Woronowicz-Peres condition) .
\item[Property 2] If $a > \frac{2}{N-1}$ then there exists a matrix in
$Int(a,N)$ which does not satisfy the Woronowicz-Peres condition .
\end{description}
It is easy to see that the extreme points of the compact convex set
$Int(a,N)$ are of the form $I +aP$ , where $P$ is an arbitrary
orthogonal projector; correspondingly $d$- dimensional vectors
composed of eigenvalues of extreme points have (up to permutations)
the following form:
$$
\lambda_{m,a} = e + aV_{m} ,  0 \le m \le d = N^{2},
$$
where $e$ is the all-ones vector and vector $V_{m}$ has its first $m$
coordinates equal to one and the rest equal to zero. Simple algebra
gives that $S(\lambda_{\frac{N(N-1)}{2}, \frac{2}{N-1}}) =1$ and 
$S(\lambda_{k,\frac{2}{N-1}}) < 1$ for all $k \neq
\frac{N(N-1)}{2}$.  Therefore if $\epsilon > 0$ is small enough then
$S(\lambda_{k,\frac{2}{N-1}+ \epsilon}) < 1$ for all $k \neq
\frac{N(N-1)}{2}$.  Corollary 2 implies that for all small enough
$\epsilon > 0$ matrices $I + (\frac{2}{N-1}+ \epsilon)P$ are are
separable (and thus satisfy the Woronowicz-Peres condition) provided
that $P$ is an orthogonal projector of rank $k \neq \frac{N(N-1)}{2}$.
It follows from Property 2 above that for all $a > \frac{2}{N-1}$
there exists an orthogonal projector $P_{a}$ such that $I +a P_{a}$
does not satisfy the Woronowicz - Peres condition, in other words that
$$
|\lambda_{min}(P_{a}^{T})|  > a^{-1} .
$$
It follows that if $a = \frac{2}{N-1}+ \epsilon$ and $\epsilon > 0$ is
small enough then necessarily
$$
Rank(P_{a}) = \frac{N(N-1)}{2}
$$ 
Thus $W_{\frac{N(N-1)}{2}}(N) \geq \frac{N-1}{2}$ , but Property 1
above implies that $W_{\frac{N(N-1)}{2}}(N) \le \frac{N-1}{2}$ .
Therefore
$$
W_{\frac{N(N-1)}{2}}(N) =  \frac{N-1}{2}.
$$

One implication of this result is 
that the boundary of the largest $p=2$ ball also contains points in the
interior of the postive cone.  One sees this by noting that 
$a = 2/(N-1)$ is the largest $a$ for which $I + a P$ is separable,
where $P$ is the rank-$N(N-1)/2$ projector achieving the value
$W_{N(N-1)/2} = (N-1)/2$ of Theorem 4.  For all greater $a$, 
the matrix is entangled; so is the scaled operator 
$(N-1)/N$ times this matrix.  But for $a= 2/(N-1)$ this 
scaled matrix satisfies
$((N-1)/N)(I + a P) = 
I + \Delta$, where $\Delta$ is Hermitian with
$N(N+1)/2$ eigenvalues
$-1/N$ and $N(N-1)/2$ eigenvalues $+1/N$.  
This is well within the interior of the positive cone, 
and $||\Delta||_2=1$.      

\section{Discussion and conclusion}

For a product of $R$ $N$-dimensional systems
in a mixture \beqa \rho = (1 - \epsilon)
I/{N^R} + \epsilon \rho'\;, \eeqa 
($\rho'$ a normalized density
matrix), \cite{Rungta2001a}, extending \cite{Braunstein99a}, 
found lower and upper bounds on the value
$\epsilon_{max}(\rho)$ below (and at) which 
the
state can be guaranteed to be separable: $1/(1 + N^{2R-1}) \le
\epsilon_{max}(\rho)  < 1/(1 + N^{R-1}$).  
(The $N=2$ case is in \cite{Braunstein99a}.)
For bipartite systems ($R=2$), these bounds are
$1/(1 + N^3)$ and $1/(1+N)$.  The lower bound is close to what one
can get from the $l_\infty$ (operator norm) result, while the 
upper bound comes from mixing in the maximally entangled $\rho_E$.  
The results of this paper give
$1/(N^2-1) \le \epsilon_{max}(\rho) \le 2/(2+N^2)$.  The lower bound is via 
Corollary 5 of Theorem 1, and is tighter due to the use of Frobenius
rather than operator norm;  the upper bound uses
Theorem 2 and the same scaling as in Cor. 5, and is tighter because
the maximally entangled state
is not the optimal state to mix in.

Our knowledge of the exact size of the 
$2$-norm ball in the bipartite
case, gives us a bound exponentially better than known bounds
on $\epsilon_{max}(\rho)$.  This shows how much more
powerful our sufficient
condition for separability is than previously known geometric conditions, 
in the bipartite setting.  

It is also illuminating to investigate the implications of our results
in the multipartite setting;  we will compare with the results 
of \cite{Rungta2001a,Braunstein99a} mentioned above. 
For multipartite states ($R>2$) we get a
slightly better upper bound on $\epsilon_{max}(\rho)$. 
For example, for even $R$,
$\epsilon_{max}(\rho) \le 2/(2 + N^R)$, 
by dividing the systems into equal
sized sets and viewing the state as bipartite.  For qubits,
this is actually the same as in \cite{Braunstein99a}, although
since we used a slightly less than optimal scaling, we could improve
it a bit.  Our results also imply that no matter what state $\rho$ is
mixed in, if $\epsilon \le 1/(N^R-1)$, the state is separable with 
respect to every bipartition.  This is dramatically larger than 
\cite{Braunstein99a}'s bound below which the state is guaranteed
separable, but not directly comparable because a state can be
separable with respect to every such bipartition
yet not be separable \cite{Bennett99a}.  This raises the important
question of the size of the largest separable ball in the 
multipartite case, to which we expect our methods can contribute.

In conclusion, we have found the exact size of the largest 
$p$-norm balls of entangled states around the identity, for 
all $1 \le p \le \infty$, and established for $p=2$
that the edge of the ball can be reached within the positive cone.
Applied via scaling as we illustrated with several corollaries and
examples, this yields sufficient conditions for separability which can be
exponentially stronger in many situations than previously known conditions.
In particular, we found the strongest such condition statable in 
terms of the spectrum of a density matrix, and derivable via scaling
of the $p=2$ result:  for normalized density matrices of $d \otimes d$
systems, it is that the purity $\tr \rho^2$ be less than $1/(d-1)$.
In addition, for three special classes of perturbations (positive 
multiples of pure states, positive multiples of the maximally
entangled state, and positive multiples of projectors), 
we found the smallest perturbation in the class achieving entanglement.
The pure state result, that it is a $2 \otimes 2$ Bell state rather
than an $N \otimes N$  
maximally entangled state, is not only mathematically interesting
but transparent in meaning, and possibly surprising.
These are natural special classes of perturbations that have been 
previously considered in quantum information theory, so we expect 
that these results
will find application in many appropriate situations.
Because of the natural geometric form of our general 
sufficient conditions for separability (Theorem 1) and related
results,
their status as a basic aspect of the geometry of the entangled
states, and the important role of these balls in computational questions,
we anticipate many applications for them, in theory and in the interpretation 
and engineering of experiments that aim to produce entanglement.

Thanks to Manny Knill for discussions, and the US DOE and NSA 
for support.

\end{document}